\documentclass[11pt,showpacs,preprintnumbers,amsmath,amssymb,prd,nofootinbib,superscriptaddress]{revtex4-2}

\usepackage{dcolumn}% Align table columns on decimal point
\usepackage{bm}% bold math
\usepackage{ifpdf}
\usepackage{hyperref}
\usepackage{bm}
\usepackage{xcolor,color,graphicx,graphics,physics}
\usepackage[spanish,english]{babel}%, portuguese
\usepackage[latin1]{inputenc}
\usepackage[OT1]{fontenc}
\usepackage{latexsym,amssymb,amsmath,amsfonts, slashed}
\usepackage{makeidx}
\usepackage{epsfig,subfigure}
\usepackage{natbib}
\usepackage{epstopdf}
\usepackage{mathrsfs}
\usepackage{hyperref}%\usepackage[colorlinks=true,linkcolor=blue]{hyperref}
\hypersetup{colorlinks=true, linkcolor=blue, citecolor=blue, urlcolor=blue}
\usepackage{enumerate}

\usepackage{fixmath}

\everymath{\displaystyle}
\usepackage{graphicx}

\usepackage[T1]{fontenc}
\usepackage{amsmath}
\usepackage{amssymb}
\usepackage{graphicx}
\usepackage{xcolor}

\newcommand{\bea}{\begin{eqnarray}}
\newcommand{\eea}{\end{eqnarray}}

\newcommand{\orcid}[1]{\href{https://orcid.org/#1}{\includegraphics[width=10pt]{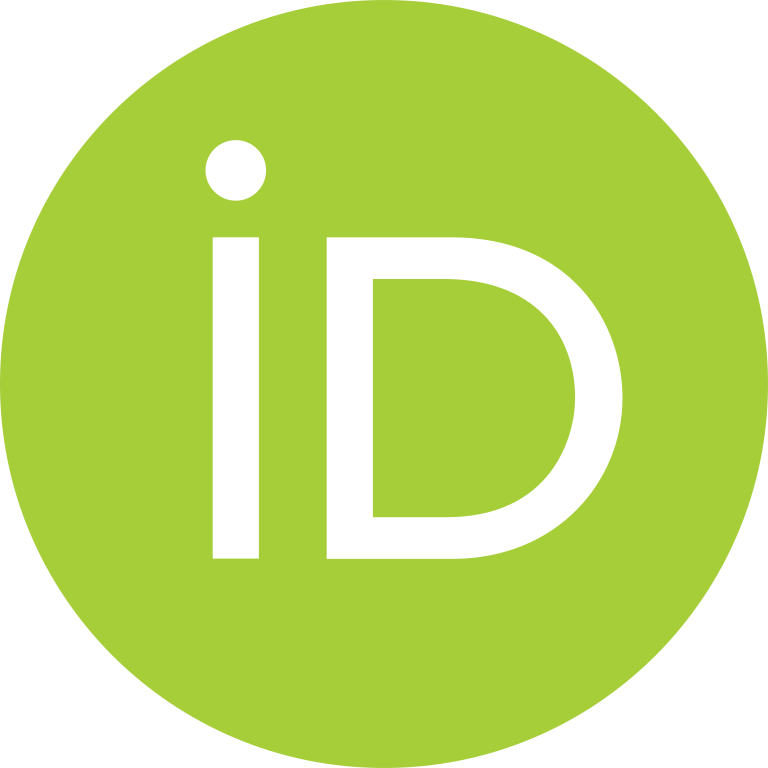}}}

\begin{document}

\title{Compton scattering in TFD formalism}

\author{D. S. Cabral  \orcid{0000-0002-7086-5582}}
\email{danielcabral@fisica.ufmt.br}
\affiliation{Instituto de F\'{\i}sica, Universidade Federal de Mato Grosso,\\
78060-900, Cuiab\'{a}, Mato Grosso, Brazil}

\author{A. F. Santos \orcid{0000-0002-2505-5273}}
\email{alesandroferreira@fisica.ufmt.br}
\affiliation{Instituto de F\'{\i}sica, Universidade Federal de Mato Grosso,\\
78060-900, Cuiab\'{a}, Mato Grosso, Brazil}

\begin{abstract}

In this paper, the cross section for the Compton scattering process at finite temperature is calculated. Temperature effects are introduced using the Thermofield Dynamics (TFD) formalism. It is a real-time finite temperature quantum field theory. Our result shows that thermal effects become relevant as the temperature increases. A comparison between the TFD and closed-time path results is presented.

\end{abstract}

\maketitle

\section{Introduction}

Compton scattering is a very important process with regard to Quantum Electrodynamics (QED), mainly because it brings with it a historical fact: one of the particular phenomena that brought about the birth of quantum mechanics \cite{comptonwoo}. However, despite being a topic discussed in all studies of basic physics, its formulation is made, in almost all cases, from the classical fundamentals to finite temperature. To make a more accurate and complete description of this scattering process, it is necessary to use some QED concepts, starting from the theory of the Standard Model \cite{gaillard,novaes,moreira}. Furthermore, it is inherent to the environment that temperature and, consequently, thermal effects on physical systems are fundamental and natural characteristics presents in any real physical process. Thus, it is necessary and interesting to analyze the Compton scattering subject to finite temperature, looking at the consequences and for the asymptotic cases.

To discuss quantum field theory at finite temperature, in particular QED theory, there are some formalisms that introduce these thermal effects into the usual zero-temperature theories. These formalisms are based on ideas that use real-time or imaginary-time transformations as a basis \cite{finite1,finite3, realandimaginary}. For example, Matsubara approach is an imaginary-time formalism, which is based on replacing the time, $t$, with a complex time, $i\tau$ \cite{Matsubara,finite2}. In this formalism, the propagators have a non-trivial dependence on temperature. For the real-time formalism there are two approaches: the closed-time path formalism \cite{Schwinger} and Thermofield Dynamics (TFD) formalism \cite{tfd1,khannatfd,Umezawa:1982nv,Umezawa:1993yq,lietfd}. The closed-time path formalism acts on the complex-time plane in such a way that the contour goes along the real axis and then returns. The TFD formalism, on the other hand, consists of two ingredients: the doubling of the Fock space and the Bogoliubov transformation. Here the TFD formalism is chosen, as its procedures and calculations are analogous to those performed in zero-temperature theories. 

The difficulty encountered in these formalisms is the fact that the relevant thermal effects, to be observed experimentally, are expected to arise when the temperatures involved are of an astronomical order, as in the center of stars, for example. Although there are several experimental results involving Compton scattering \cite{experimental1,experimental2,experimental3,experimental4}, these measurements are obtained in laboratories in which the experiments are  carried out as close as possible to the ideal situation of zero temperature. Based on this fact, the present work describes the Compton scattering at finite temperature, using the QED formalism and introducing the thermal effects via TFD formalism. Although the temperature effects for Compton scattering have already been calculated using the closed-time path formalism \cite{comptonclosed-time}, in this paper such a study is developed following a different approach. It is important to note that TFD and closed-time path are real-time approaches with different procedures and tools. There are similarities between these formalisms, however, these analogies should not go too far. The structure of the propagator in the closed-time path formalism and in the TFD formalism is not the same. In TFD the fields of the duplicated Hilbert space commute or anti-commute, while in the closed-time formalism the operators on the path above the real axis do not commute or anti-commute with those on the path below the real axis \cite{Umezawa:1993yq, Das, Ahmed}. Here the result for a scattering process is considered and it is compared in  both approaches. The differential cross section at high temperatures, at least for the tree-level, exhibits different behavior.

This paper is organized as follows. In section II, a brief introduction to the TFD formalism is presented. In section III, Compton scattering process is introduced.  Feynman diagrams characterizing this scattering are shown. The probability amplitude at finite temperature is obtained. In section IV, the differential cross section for this problem is calculated. The usual result at zero temperature is recovered at the proper limit. Our result is compared with solutions obtained from the closed-time path approach. Finally, in the Section V, some final considerations are discussed.

\section{Thermofield Dynamics} 

TFD is a formalism that introduces temperature into quantum fields from ideas of symmetries and Lie groups, based on concepts of statistical mechanics and thermodynamics \cite{khannatfd,Umezawa:1982nv,Umezawa:1993yq}. It is done by doubling the Hilbert space so that from this doubling, one can write a thermal Hilbert space $\mathbb{H}_T=\mathbb{H}\otimes\widetilde{\mathbb{H}}$, where $\mathbb{H}$ deals with the usual Hilbert space, while $\widetilde{\mathbb{H}}$ is obtained from the former by the tilde conjugation rules. For a dynamical operator $A$, these rules are defined as
\begin{equation}
\begin{split}
        (A_iA_j)^{\widetilde{}}&=\widetilde{A}_i\widetilde{A}_j;\\
        (cA_i+A_j)^{\widetilde{}}&=c^{*}\widetilde{A}_i+\widetilde{A}_j;\\
        (A_i^{\dagger})^{\widetilde{}}&=(\widetilde{A}_i)^{\dagger};\\
        (\widetilde{A}_i)^{\widetilde{}}&=\pm A_i;\\
        [A_i,\Tilde{A}_j]&=0.
    \end{split}
\end{equation}
Here the positive (negative) sign represents the bosonic (fermionic) quantization. Note that $A$ acts on $\mathbb{H}$ while $\widetilde{A}$ acts on $\widetilde{\mathbb{H}}$.

In this approach, the expected value of a dynamical operator $A$ is given by $\langle A\rangle=\bra{0(\beta)}A\ket{0(\beta)}$, where $\ket{0(\beta)}$ is the thermal vacuum state defined as   
\begin{equation}
\ket{0(\beta)}=\frac{1}{\sqrt{Z(\beta)}}\sum_ne^{-\frac{1}{2}\beta E_n}\ket{n,\widetilde{n}},
\end{equation}
and $Z(\beta)$ the canonical partition function, with $\beta=1/k_B T$, being $k_B$ the Boltzmann constant and $T$ the temperature. The state $\ket{n,\widetilde{n}}$ is in the thermal space $\mathbb{H}_T$, such that a generic state in it can be written as
\begin{equation}
\ket{m,\widetilde{n}}=\frac{1}{\sqrt{m!}\sqrt{n!}}(a^{\dagger})^m(\widetilde{a}^{\dagger})^n\ket{0,\tilde{0}}
\end{equation}
with
$\ket{0,\tilde{0}}=\ket{0}\otimes\ket{\tilde{0}}$, where this first is the vacuum state of $\mathbb{H}$, while the second one has a similar role in $\widetilde{\mathbb{H}}$. In this approach, the operators $a^{\dagger}$ and $\widetilde{a}^{\dagger}$, along with $a$ and $\widetilde{a}$ have the role of creation and annihilation operators in their respective spaces.

Furthermore, this formalism also brings the Bogoliubov transformations. These are unitary and act by connecting thermal and non-thermal operators through a rotation in the double space. This implies that the thermal creation and annihilation operators act on $\mathbb{H}_T$ as
\begin{equation}
\ket{m(\beta),\tilde{n}(\beta)}=\frac{1}{\sqrt{m!}}\frac{1}{\sqrt{n!}}(a^{\dagger}_\beta)^m(\tilde{a}^{\dagger}_\beta)^n\ket{0(\beta)},
\end{equation}
where the $\beta$ index represents the temperature dependence of the operators.

For fermions, whose creation and annihilation operators are $a^{\dagger}(p)$ and $a(p)$, respectively, there are the following transformations,
\begin{equation}
a^\dagger(p)=U(\beta)a_\beta^\dagger(p)+V(\beta)\tilde{a}_\beta(p);\quad\quad a(p)=U(\beta)a_\beta(p)+V(\beta)\tilde{a}_\beta^\dagger(p),\label{eq10}
\end{equation}
where
\begin{equation}
    U=\frac{1}{\sqrt{1+e^{-\beta E}}};\quad\quad V=\frac{1}{\sqrt{1+e^{\beta E}}}.\label{eq02}
\end{equation}
The bosons follow similar thermal transformations, i.e.
\begin{equation}
d^\dagger(p)=U^{\prime}(\beta)d_\beta^\dagger(k)+V^{\prime}(\beta)\tilde{d}_\beta(k);\quad\quad d(p)=U^{\prime}(\beta)d_\beta(k)+V^{\prime}(\beta)\tilde{d}_\beta^\dagger(k), \label{eq11}
\end{equation}
in such a way that
\begin{equation}
        U^{\prime}=\frac{1}{\sqrt{1-e^{-\beta E}}};\quad\quad V^{\prime}=\frac{e^{-\beta E/2}}{\sqrt{1-e^{-\beta E}}}.\label{eq03}
\end{equation}

With these definitions, the thermal operators of creation and annihilation of bosons (fermions) obey the following commutation (anti-commutation) relations in equal time
\begin{equation}
[d_\beta(k,\lambda),d_\beta^{\dagger}(k^{\prime},\lambda^{\prime})]=\frac{1}{N_{k}}\delta_{\lambda,\lambda^{\prime}}\delta^3(\vec{k}-\vec{k}^\prime),\quad\quad\{a_\beta(p,s),a_\beta^{\dagger}(p^{\prime},s^{\prime})\}=\frac{1}{N_{p}}\delta_{s,s^{\prime}}\delta^3(\vec{p}-\vec{p}^\prime),
\end{equation}
with $N_p$ and $N_k$ being normalization factors.

The formalism presented here will be used in the next sections to calculate the differential cross section for Compton scattering at finite temperature.

\section{Compton scattering}

In this section, the main objective is to calculate the probability amplitude at finite temperature for Compton scattering. The Lagrangian that describes this process is given as
\begin{equation}
    \mathcal{L}=-\frac{1}{4}F_{\mu\nu}F^{\mu\nu}+\bar{\psi}\left(i\gamma^\mu D_\mu-m\right)\psi,
\end{equation}
where $F_{\mu\nu}=\partial_\mu A_\nu-\partial_\nu A_\mu$ is the electromagnetic tensor, $\psi$ is the fermion field, $\gamma^\mu$ are the Dirac matrices, $m$ is the fermion mass and $D_\mu=\partial_\mu-ieA_\mu$ with $e$ being the electron charge and $A_\mu$ the photon field. For the interest proposed here, the main part is the interaction Lagrangian written as
\begin{equation}
    \mathcal{L}_{\text{int}}=e\bar{\psi}\gamma^\mu\psi A_\mu.
\end{equation}
As a consequence, the interaction Hamiltonian is
\begin{equation}
    \mathcal{H}_{\text{int}}=-\mathcal{L}_{\text{int}}=-e\bar{\psi}\gamma^\mu\psi A_\mu.
\end{equation}

To deal with a theory at finite temperature, the workspace is doubled, i.e. generating the tilde space. And from this, one can write the operators that generate the symmetry, as the $\hat{\mathcal{H}}$ which is related to the time evolution. This operator is defined as 
\begin{equation}
    \hat{\mathcal{H}}=\mathcal{H}-\tilde{\mathcal{H}},
\end{equation}
where $\mathcal{H}$ and $\tilde{\mathcal{H}}$ are operators in standard Hilbert space and tilde space, respectively.

\begin{figure}
    \centering
    \includegraphics[scale=0.5]{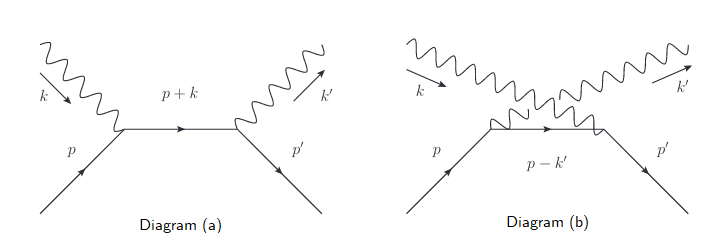}
\caption{Feynman diagrams of the Compton scattering. Adapted from Ref. \cite{compton}.}
    \label{fig1}
\end{figure}

Compton scattering is the reaction between an electron and a photon through a virtual electron, according to the Feynman diagrams shown in Figure \ref{fig1}.
For this process, in the TFD formalism, the initial and final states are
\begin{equation}
\ket{i}=\ket{e_{i},\gamma_i}=a^{\dagger}_\beta(p)d^{\dagger}_\beta(k)\ket{0(\beta)};\quad\quad\ket{f}=\ket{e_f,\gamma_f}=a^{\dagger}_\beta(p^{\prime})d^{\dagger}_\beta(k^{\prime})\ket{0(\beta)},
\end{equation}
where $e_i$ ($e_f$) and $\gamma_i$ ($\gamma_f$) represent the initial (final) states of the electron and photon, respectively.

In quantum electrodynamics, the second order term of the scattering matrix is given by
\begin{equation}
\hat{S}^{(2)}=\frac{(-i)^2}{2}\int d^4xd^4y:\hat{\mathcal{H}}_{\text{int}}(x)\hat{\mathcal{H}}_{\text{int}}(y):\;\;=\frac{(-i)^2}{2}\int d^4xd^4y:\mathcal{H}_{\text{int}}(x)\mathcal{H}_{\text{int}}(y):,\label{eq09}
\end{equation}
with $:(\cdots):$ being the normal ordering operator, which introduces the correct order for the operators to be applied, in order to avoid divergences. In addition, to the last term in (\ref{eq09}), only the non-tilde operators are considered, since these are actually the observables of the theory.

For this scattering process, the probability amplitude is
\begin{equation}
\mathcal{M}=\bra{f}\hat{S}^{(2)}\ket{i}=-\frac{e^2}{2}\int d^4xd^4y\bra{f}:\bar{\psi}(x)\gamma^\mu\psi(x)\bar{\psi}(y)\gamma^\nu\psi(y)A_{\mu}(x)A_\nu(y):\ket{i},
\end{equation}
so that, after an analysis of the field operators, it can be seen that only the terms described by the Feynman diagrams will contribute to the scattering. Then
\begin{equation}
    \begin{split}
    \mathcal{M}&=-e^2\int d^4xd^4y\bra{f}:\Bar{\psi}^{-}(x)\gamma^\mu[\psi(x)\Bar{\psi}(y)]\gamma^\nu\psi^{+}(y)A_\mu^{-}(x)A_\nu^{+}(y):\ket{i}\\&-e^2\int d^4xd^4y\bra{f}:\Bar{\psi}^{-}(x)\gamma^\mu[\psi(x)\Bar{\psi}(y)]\gamma^\nu\psi^{+}(y)A_\mu^{+}(x)A_\nu^{-}(y):\ket{i}\equiv \mathcal{M}_a+\mathcal{M}_b,\label{eq01}
    \end{split}
\end{equation}
where $\mathcal{M}_a$, $\mathcal{M}_b$ corresponding to the diagrams $(a)$ and $(b)$ in Figure \ref{fig1}, respectively, and the positive (negative) index represents the part of field operators with positive (negative) frequencies, in other words
\begin{equation}
    \psi^{+}(x)=\int d^3pN_p\sum_sa(p,s)u(p,s)e^{-ipx};\quad\quad \bar{\psi}^{-}(x)=\int d^3pN_p\sum_sa^{\dagger}(p,s)\bar{u}(p,s)e^{ipx}
\end{equation}
for the fermionic field, and
\begin{equation}
A^{+}_\mu(x)=\int d^3kM_k\sum_\lambda d(k,\lambda)\epsilon_\mu(k,\lambda)e^{-ikx};\quad\quad
    A^{-}_\mu(x)=\int d^3kM_k\sum_\lambda d^{\dagger}(k,\lambda)\epsilon^{\dagger}_\mu(k,\lambda)e^{ikx},
\end{equation}
for the photonic field. Note that the four-vector product $px=\xi t-\vec{p}\cdot\vec{x}$ leads to $\xi^2=\vec{p}^2+m^2$, where $\xi$ is the energy and $\vec{p}$ is the 3-momentum.

Therefore, from Eq. (\ref{eq01}) suppressing the spin and polarization indices, taking into account the Bogoliubov transformations  Eq. (\ref{eq10}) and Eq. (\ref{eq11}) for the operators, the probability amplitude to the first diagram becomes
\begin{equation}
    \begin{split}
    \mathcal{M}_a&=-e^2\int d^4xd^4y\int\frac{d^4q}{(2\pi)^4}(U^{\prime})^2U^2\left[\bar{u}(p^{\prime})\gamma^\mu e^{i(p^{\prime}+k^{\prime}-q)x}S(q)e^{-i(p+k-q)y}\gamma^\nu u(p)\epsilon_\mu^{\dagger}(k^{\prime})\epsilon_\nu(k)\right],
    \end{split}
\end{equation}
where $S(q)=S^{(0)}(q)+S^{(\beta)}(q)$ is the electron propagator in the momentum space, with
\begin{equation}
    S^{(0)}(q)=\frac{\slashed{q}+m}{q^2-m^2},
\end{equation}
being the zero temperature part, and
\begin{equation}
    S^{(\beta)}(q)=\frac{\pi i}{(e^{\beta q^0}+1)}\left[\frac{(\gamma^0\xi-\gamma^{i}q_i+m)}{2\xi}\Delta_1\delta(q^0-\xi)+\frac{(\gamma^0\xi+\gamma^{i}q_i-m)}{2\xi}\Delta_2\delta(q^0+\xi)\right]
\end{equation}
is the temperature dependent part \cite{ale1}. The terms $\Delta_1$ and $\Delta_2$ arise from the doubled space and are thermal matrices $2\times2$ given by
\begin{equation}
    \Delta_1=\begin{pmatrix}
    1 & e^{\beta q_0/2}\\e^{\beta q_0/2}&-1\end{pmatrix};\quad\quad \Delta_2=\begin{pmatrix}
    -1 & e^{\beta q_0/2}\\e^{\beta q_0/2}&1\end{pmatrix}.
\end{equation}

From the thermal functions Eqs. (\ref{eq02}) and (\ref{eq03}), derived from the Bogoliubov transformations, we get
\begin{equation}
(U^\prime)^2U^2=\frac{1}{1-e^{-2\beta E}}=\frac{1+\coth\beta E}{2}
\end{equation}
and using the definition of the Dirac delta function
\begin{equation}
    \int d^4xd^4ye^{i(p^{\prime}+k^{\prime}-q)x}e^{-i(p+k-q)y}=\delta^4(p^{\prime}+k^{\prime}-q)\delta^4(p+k-q),
\end{equation}
the probability amplitude for the diagram $(a)$ becomes
\begin{equation}
    \mathcal{M}_a=-e^2\int\frac{d^4q}{(2\pi)^4}\left[\bar{u}(p^{\prime})\gamma^\mu \delta^4(p^{\prime}+k^{\prime}-q)S(q)\delta^4(p+k-q)\gamma^\nu u(p)\epsilon_\mu^{\dagger}(k^{\prime})\epsilon_\nu(k)\right]\left(\frac{1+\coth{\beta E}}{2}\right),
\end{equation}
in such a way that, performing the integration over the momentum space of the electron in the middle, remembering that the Dirac delta function imposes momentum conservation, we obtain
\begin{equation}
    \mathcal{M}_a=-e^2\bar{u}(p^{\prime})\gamma^\mu S(p+k)\gamma^\nu u(p)\epsilon_\mu^{\dagger}(k^{\prime})\epsilon_\nu(k)\frac{1+\coth{\beta E}}{2}.
\end{equation}
Analogously, for the diagram $(b)$ the probability amplitude is
\begin{equation}
    \mathcal{M}_b=-e^2 \bar{u}(p^{\prime})\gamma^\mu S(p-k^{\prime})\gamma^\nu u(p)\epsilon_\mu(k)\epsilon_\nu^{\dagger}(k^{\prime})\frac{1+\coth{\beta E}}{2}.
\end{equation}

In the next section, these results are used to calculate the differential cross section for the Compton scattering at finite temperature in the TFD approach.

\section{Differential Cross section}

Here the differential cross section at finite temperature is obtained for Compton scattering. To achieve this objective, the main quantity to be calculated is the probability density. Since the scattering amplitude Eq. (\ref{eq01}) is given by $\mathcal{M}=\mathcal{M}_a+\mathcal{M}_b$, the probability density is given as
\begin{equation}
    |\mathcal{M}|^2=|\mathcal{M}_a|^2+|\mathcal{M}_b|^2+2\Re{\mathcal{M}_a^\dagger\mathcal{M}_b}.
\end{equation}
Taking into account the unpolarized initial and final beams, it is necessary to calculate the average over all spins and polarizations. Then
\begin{equation}
\langle|\mathcal{M}|^2\rangle=\frac{1}{4}\sum_{\substack{\lambda,\lambda^{\prime}\\s,s^{\prime}}}|\mathcal{M}|^2=\langle|\mathcal{M}_a|^2\rangle+\langle|\mathcal{M}_b|^2\rangle+\left\langle2\Re{\mathcal{M}_a^\dagger\mathcal{M}_b}\right\rangle.\label{30}
\end{equation}

The average of the diagram $(a)$ leads to
\begin{eqnarray}
\langle|\mathcal{M}_a|^2\rangle  &=&\frac{1}{4}\sum_{\substack{\lambda,\lambda^{\prime}\\s,s^{\prime}}}e^4\left[\bar{u}(p^{\prime})\gamma^\mu S(p+k)\gamma^\nu u(p)\epsilon_\mu^{\dagger}(k^{\prime})\epsilon_\nu(k)\right]\nonumber\\
&&\times\left[\epsilon^{\dagger}_\omega(k)\epsilon_\alpha(k^{\prime})\bar{u}(p)\gamma^\omega S^{\dagger}(p+k)\gamma^\alpha u(p^{\prime})\right]
\left(\frac{1+\coth{\beta E}}{2}\right)^2.\label{31}
\end{eqnarray}
Using the completeness relations
\begin{equation}
\sum_{s}u(p,s)\bar{u}(p,s)={\slashed{p}+m};\quad\quad\sum_{\lambda}\epsilon_\mu(k,\lambda)\epsilon^{\dagger}_\alpha(k,\lambda)=g_{\mu\alpha},
\end{equation}
and the relation
\begin{equation}
[\bar{u}(p)\gamma^\mu\gamma^\nu u(p^\prime)\bar{u}(p^\prime)\gamma_\mu\gamma_\nu u(p)]=\Tr[u(p)\bar{u}(p)\gamma^\mu\gamma^\nu u(p^\prime)\bar{u}(p^\prime)\gamma_\mu\gamma_\nu],
\end{equation}
as well as the following property of Dirac matrices
\begin{equation}
\gamma^\mu\slashed{p}\gamma_\mu=-2\slashed{p},
\end{equation}
Eq. (\ref{31}) reads
\begin{equation}
\langle|\mathcal{M}_a|^2\rangle ={e^4}\Tr{\slashed{p}^{\prime}S(p+k)\slashed{p}S^{\dagger}(p+k)}\left(\frac{1+\coth{\beta E}}{2}\right)^2.
\end{equation}
Splitting the propagator into its zero and finite temperature parts, we have
\begin{eqnarray}
\langle|\mathcal{M}_a|^2\rangle&=&{e^4}\left\{\Tr[\slashed{p}^{\prime}S^{(0)}(p+k)\slashed{p}S^{(0)}(p+k)]+\Tr[\slashed{p}^{\prime}S^{(0)}(p+k)\slashed{p}S^{(\beta)\dagger}(p+k)]\right.\nonumber\\
&&\left.+\Tr[\slashed{p}^{\prime}S^{(\beta)}(p+k)\slashed{p}S^{(0)}(p+k)]+\Tr[\slashed{p}^{\prime}S^{(\beta)}(p+k)\slashed{p}S^{(\beta)\dagger}(p+k)]\right\}\nonumber\\
&&\times\left(\frac{1+\coth{\beta E}}{2}\right)^2.\label{eq04}
\end{eqnarray}

Considering the centre of mass frame, in which the four momenta are given by
\begin{eqnarray}
    p&=&(E,0,0,-\omega);\quad k=(\omega,0,0,\omega);\nonumber\\
     p^{\prime}&=&(E,-\omega\sin{\theta},0,-\omega\cos{\theta});\quad k^{\prime}=(\omega,\omega\sin{\theta},0,\omega\cos{\theta}),
\end{eqnarray}
with $\theta$ being the scattering angle, and also the high energy approximation $m\to0$ (which implies $E\to\omega$), we get
\begin{equation}
\Tr[\slashed{p}^{\prime} S^{(0)}(p+k)\slashed{p} S^{(0)}(p+k)]=\frac{8}{(p+k)^4}(p^{\prime}\cdot k)(p\cdot k).
\end{equation}
The same result has been obtained by \cite{compton}, for the analysis of the scattering at zero temperature.

The other terms of Eq. (\ref{eq04}) are all null, because for $p+k=(E+\omega,0,0,0)$, in the propagator at the centre of mass frame subject to high energies approximation, we have $(p+k)_i=0$ and, consequently, $\xi=0$. Therefore, the average over the first diagram is
\begin{equation}
\begin{split}
\langle|\mathcal{M}_a|^2\rangle  &=e^4(\cos{\theta}+1)\left(\frac{1+\coth{\beta E}}{2}\right)^2.
\end{split}\label{eq05}
\end{equation}

In a similar way, the average of the diagram $(b)$, where the momentum conservation is $p-k^\prime=(E-\omega,\vec{p}+\vec{p}^{\prime})$, is
\begin{eqnarray}
  \langle|\mathcal{M}_b|^2\rangle  &=&{e^4}\left\{\Tr[\slashed{p}^{\prime}S^{(0)}(p-k^\prime)\slashed{p}S^{(0)}(p-k^\prime)]+\Tr[\slashed{p}^{\prime}S^{(0)}(p-k^\prime)\slashed{p}S^{(\beta)\dagger}(p-k^\prime)]\right.\nonumber\\
  &&\left.+\Tr[\slashed{p}^{\prime}S^{(\beta)}(p-k^\prime)\slashed{p}S^{(0)}(p-k^\prime)]+\Tr[\slashed{p}^{\prime}S^{(\beta)}(p-k^\prime)\slashed{p}S^{(\beta)\dagger}(p-k^\prime)]\right\}\nonumber\\
  &&\times\left(\frac{1+\coth{\beta E}}{2}\right)^2,\label{eq06}
\end{eqnarray}
where
\begin{equation}
    \Tr[\slashed{p}^\prime S^{(0)}(p-k^\prime)\slashed{p} S^{(0)}(p-k^\prime)]=\frac{8}{(p-k^\prime)^4}(p^{\prime}\cdot k^{\prime})(k^{\prime}\cdot p),
\end{equation}
as in the case of zero temperature \cite{compton}.

The next terms in Eq. (\ref{eq06}) imply that
\begin{equation}
\Tr[\slashed{p}^{\prime}S^{(0)}(p-k^\prime)\slashed{p}S^{(\beta)\dagger}(p-k^\prime)]=-\Tr[\slashed{p}^{\prime}S^{(\beta)}(p-k^\prime)\slashed{p}S^{(0)}(p-k^\prime)],
\end{equation}
and, finally
\begin{eqnarray}
\Tr[\slashed{p}^{\prime}S^{(\beta)}(p-k^\prime)\slashed{p}S^{(\beta)\dagger}(p-k^\prime)]&=&\frac{8\pi^2\omega^2}{(e^{\beta q^0}+1)^24\xi_b^2}\Bigl\{\cos{\left(\frac{\theta}{2}\right)}\bigl[(2\omega-\xi_b)^2\delta^2(q^0-\xi_b)\nonumber\\
&&+(2\omega+\xi_b)^2\delta^2(q^0+\xi_b)\bigl]\nonumber\\
&&+\left[4\omega^2(\cos{\theta}+1)-2\xi_b^2\cos{\theta}\right]\delta(q^0-\xi_b)\delta(q^0+\xi_b)\Bigl\},
\end{eqnarray}
with $q_0=E-\omega$, $\xi_b=2\omega\cos{(\theta/2)}$ and $E\to\omega$ at the high energy limit. In such a way that
\begin{eqnarray}
    \langle|\mathcal{M}_b|^2\rangle  &=&\Biggl\{\frac{4e^4}{\cos{\theta}+1}+
\frac{\pi^2e^2\omega^2}{2[e^{\beta(E-\omega)}+1]^2}\sec{\left(\frac{\theta}{2}\right)}\left[10\cos{\theta}+\cos{2\theta}+17\right]\nonumber\\
&&\times\delta^2\left(2\omega\cos{\frac{\theta}{2}}\right)\Biggl\}\left(\frac{1+\coth{\beta E}}{2}\right)^2,
\label{eq14}
\end{eqnarray}
where only the first element of the resulting thermal matrix was taken, since only terms of the non-tilde space are related to the observables. For the mixing term, we get
\begin{equation}
\left\langle2\Re{\mathcal{M}_a^\dagger\mathcal{M}_b}\right\rangle=0.\label{eq07}
\end{equation}
The Mandelstam variables have been used
\begin{equation}
   s=2p\cdot k=2p^\prime\cdot k^\prime;\quad\quad t=-2p\cdot p^\prime=-2k\cdot k^\prime;\quad\quad u=-2p\cdot k^\prime=-2p^\prime\cdot k;
\end{equation}
or in terms of the scattering angle of the centre of mass frame 
\begin{equation}
     s=(2\omega)^2;\quad\quad t=-2\omega^2(1-\cos{\theta});\quad\quad u=-2\omega^2(1+\cos{\theta}).
\end{equation}

With Eqs. (\ref{eq05}), (\ref{eq14}) and (\ref{eq07}), Eq. (\ref{30}) becomes
\begin{equation}
\begin{split}
\langle|\mathcal{M}|^2\rangle&=2e^4\left(\frac{\cos{\theta}+1}{2}+\frac{2}{\cos{\theta}+1}\right)\left(\frac{1+\coth{\beta E}}{2}\right)^2\\&+
\frac{\pi^2e^2\omega^2\sec{\left(\frac{\theta}{2}\right)}\left[10\cos{\theta}+\cos{2\theta}+17\right]}{2[e^{\beta(E-\omega)}+1]^2}\delta^2\left(2\omega\cos{\frac{\theta}{2}}\right)\left(\frac{1+\coth{\beta E}}{2}\right)^2.
\end{split}
\end{equation}

%It is necessary to be careful with the exponential in the second term, since the high energies approximation implies that $E\to\omega$. In other words, one can take $q_0=0$ in the delta functions, but not in this term, because the $\beta$ may come to diverge. In this case the product $\beta(E-\omega)$ would be undefined. To avoid this problem, is considered $(E-\omega)\to0$ but $E\neq\omega$.

Thus, using the definition of the differential cross section for any scattering process, i.e.,
\begin{equation}
    \frac{d\sigma}{d\Omega}=\frac{1}{64\pi^2s}\langle|\mathcal{M}|^2\rangle,
\end{equation}
for Compton scattering at finite temperature, we get
\begin{equation}
\begin{split}
\left(\frac{d\sigma}{d\Omega}\right)_\beta&=\frac{2e^4}{64\pi^2(2\omega)^2}\left(\frac{\cos{\theta}+1}{2}+\frac{2}{\cos{\theta}+1}\right)\left(\frac{1+\coth{\beta E}}{2}\right)^2\\&+\frac{\pi^2e^2\omega^2\sec{\left(\frac{\theta}{2}\right)}\left[10\cos{\theta}+\cos{2\theta}+17\right]}{64\pi^2(2\omega)^22[e^{\beta(E-\omega)}+1]^2}\delta^2\left(2\omega\cos{\frac{\theta}{2}}\right)\left(\frac{1+\coth{\beta E}}{2}\right)^2.\label{eq08}
\end{split}
\end{equation}
It should be noted that for the low temperature limit $T\to0$, it leads to $\beta\to\infty$, such that
\begin{equation}
\lim_{\beta\to\infty}\coth{\beta E}=1;\quad\quad\lim_{\beta\to\infty}\frac{1}{[e^{\beta(E-\omega)}+1]^2}=0,
\end{equation}
therefore
\begin{equation}
\lim_{T\to0}\left(\frac{d\sigma}{d\Omega}\right)_\beta=\frac{2e^4}{64\pi^2(2\omega)^2}\left(\frac{\cos{\theta}+1}{2}+\frac{2}{\cos{\theta}+1}\right)=\left(\frac{d\sigma}{d\Omega}\right).
\end{equation}
This is the differential cross section for Compton scattering at zero temperature, as obtained by \cite{compton}.

For the high temperatures limit, $\beta\to0$, we have $[e^{\beta(E-\omega)}+1]^2\to1$, but $\coth{\beta}\to\infty$. Then the temperature contribution becomes a leading term and the thermal effects are more evident and relevant. Furthermore, it has no problem dealing with a product of delta functions with identical arguments. In other words, there are regularized forms to work with these functions. Which means that their values exists and are finite \cite{realandimaginary}.

In terms of the differential cross section at zero temperature, one can rewrite Eq. (\ref{eq08}) as
{\small
\begin{eqnarray}
        \left(\frac{d\sigma}{d\Omega}\right)_\beta&=& \left(\frac{d\sigma}{d\Omega}\right)\left[1+\frac{\pi^2\omega^2}{4[e^{\beta(E-\omega)}+1]^2}\frac{2\sec{\left(\frac{\theta}{2}\right)}\left[55\cos{\theta}+12\cos{2\theta}+\cos{3\theta}+44\right]}{4\cos{\theta}+\cos{2\theta}+11}\delta^2\left(2\omega\cos{\frac{\theta}{2}}\right)\right]\nonumber\\
   &&\times\left(\frac{1+\coth{\beta E}}{2}\right)^2.
\end{eqnarray}
}
In order to compare our result with the result obtained using the closed-time path (CTP) formalism \cite{comptonclosed-time}, for simplicity, will be considered
\begin{equation}
\delta^2\left(2\omega\cos{\frac{\theta}{2}}\right)=\frac{2}{\pi^2\omega^2}\frac{4\cos{\theta}+\cos{2\theta}+11}{\sec{\left(\frac{\theta}{2}\right)}\left[55\cos{\theta}+12\cos{2\theta}+\cos{3\theta}+44\right]}.
\end{equation}
Then, one can write $\left(\frac{d\sigma}{d\Omega}\right)_\beta=F^2_{\text{TFD}}(\beta) \left(\frac{d\sigma}{d\Omega}\right)$, with
\begin{equation}
    F^2_{\text{TFD}}(\beta)\equiv\left\{1+\frac{1}{[e^{\beta(E-\omega)}+1]^2}\right\}\left(\frac{1+\coth{\beta E}}{2}\right)^2.\label{eq12}
\end{equation}

\begin{figure}[ht]
    \centering
\includegraphics[scale=0.5]{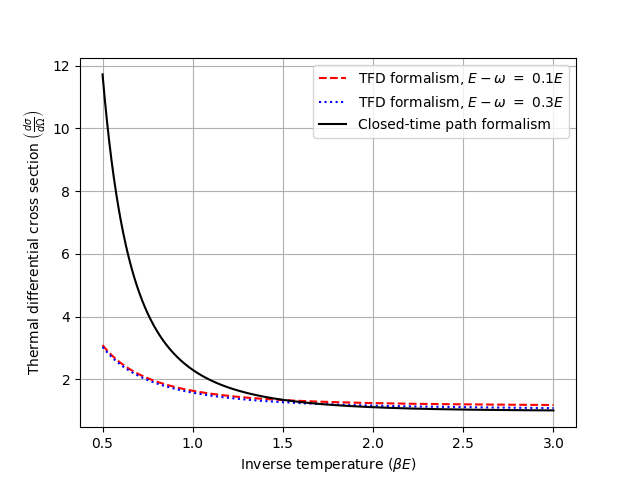}
    \caption{Thermal differential cross section in terms of the cross section at zero temperature, for the closed-time path formalism (black solid line), TFD with $E-\omega=0.1E$ (red dashed line) and TFD with $E-\omega=0.3E$ (blue dotted line). The horizontal axis is normalized by an fixed energy.}
    \label{fig2}
\end{figure}

A similar result using the closed-time path formalism was obtained by \cite{comptonclosed-time}, i.e. $\left(\frac{d\sigma}{d\Omega}\right)_\beta=F^2_{\text{CTP}}(\beta) \left(\frac{d\sigma}{d\Omega}\right)$, where
\begin{equation}
    F^2_{\text{CTP}}=\left(\frac{1+\coth{\beta E}}{2}\right)^2\coth^2{\beta E}.\label{eq13}
\end{equation}
The functions (\ref{eq12}) and (\ref{eq13}) are represented in Figure \ref{fig2}. 

It is important to note that, although both formalisms have similar behavior, the functions (\ref{eq12}) and (\ref{eq13}) are different, then at high temperature the behavior is different as shown in Figure \ref{fig2}. This difference arises because the procedures and tools used in these formalisms are completely different. The main difference is due to the form in which the average of an operator is calculated. In the TFD formalism, the main idea is to interpret the statistical average of an arbitrary operator as its expected value in a thermal vacuum $|0(\beta)\rangle$. Two elements are needed to construct this thermal state: (i) doubling of the original Hilbert space and (ii) Bogoliubov transformations. In the closed-time path approach, the statistical average is calculated using the density matrix $\rho$ \cite{comptonclosed-time}. 
Therefore, it is important to emphasize that the result obtained here uses the Bogoliubov transformation to introduce temperature effects, while in the closed-time path approach the thermal effects are introduced in a different way. In addition, as stated in reference \cite{comptonclosed-time} in the imaginary-time formalism such an analysis is not possible, since the Matsubara formalism cannot give a first principles formulation of thermal scattering because it is impossible to define the transition probability, and hence the cross-section. Then, there is no general formulation to define a thermal scattering process, although TFD and closed-time path present the same functional behavior as shown in Figure 2. For more details related to the similarities and differences between these formalisms, see the references \cite{Umezawa:1993yq, Das}.

\section{Conclusion}

It is known that there are two real-time approaches, i.e. TFD and closed-time path, to introducing temperature effects into a quantum field theory. However, solving finite temperature field theory problems using the TFD formalism gives us calculations very similar to those present in zero temperature theories. This is the main advantage over using the closed-time path formalism. The results obtained here, although not the same, are similar. It can be noticed that, for very high temperatures, i.e. very low values of $\beta$, the thermal effects will be dominant. In this case, the differential cross section at finite temperature will be many times greater than at zero temperature. Quantitatively speaking, at centre of mass energies on the order of $1KeV$, thermal effects would be detected at temperatures on the order of $10^7K$. This corresponds to situations in the center of stars, such as the Sun, for example.

From Figure \ref{fig2} it can be seen that the solution obtained by the closed-time path formalism grows faster when $\beta\to0$. Therefore, the thermal effects at high temperatures, in this approach, are more expressive than those obtained by the TFD formalism. This can be understood from Eqs. (\ref{eq12}) and (\ref{eq13}), which shows that $F^2_{\text{CTP}}>F^2_{\text{TFD}}$ when very high values of  $T$ are assumed. Still in Figure \ref{fig2}, the TFD formalism for this specific scattering strongly depends on the difference $E-\omega$, due to the presence of the exponential in the first term of (\ref{eq12}). What comes from the centre of mass frame in approaching high energies. This fact implies a slower decay in the asymptotic limit $\beta\to\infty$ of the TFD function compared to that of closed-time path. Therefore, despite the different contributions, it can be noted that both approaches have the same functional behavior.

\section*{Acknowledgments}

This work by A. F. S. is partially supported by National Council for Scientific and Technological Develo\-pment - CNPq project No. 313400/2020-2. 

%%%%%%%%%%%%%%%%%%%%%%%%%%%%%%%%%%%%%%%%%%%%%%%%%%%%%%%%%%%%%%%%%%%%%%%%%%%%%%%%%%%%%%%%%%%%%%%%%%%%%%%%%%%%%%%%%

\global\long\def\link#1#2{\href{http://eudml.org/#1}{#2}}
 \global\long\def\doi#1#2{\href{http://dx.doi.org/#1}{#2}}
 \global\long\def\arXiv#1#2{\href{http://arxiv.org/abs/#1}{arXiv:#1 [#2]}}
 \global\long\def\arXivOld#1{\href{http://arxiv.org/abs/#1}{arXiv:#1}}

%%%%%%%%%%%%%%%%%%%%%%%%%%%%%%%%%%%%%%%%%%%%%%%%%%%%%%%%%%%%%%%%%%%%%%%%%%%%%%%%%%%%%%%%%%%%%%%%%%%%%%%%%%%


\begin{thebibliography}{99}

\bibitem{comptonwoo} Y. H. Woo, ``The compton effect and tertiary x-radiation,'' 
\doi{} {Proceedings of the National Academy of Sciences of the United States of America {\bf 11}, 123 (1925).}

\bibitem{gaillard} M. K. Gaillard, P. D. Grannis, and F. J. Sciulli, ``The standard model of particle physics,'' 
\doi{10.1103/RevModPhys.71.S96} {Rev. Mod. Phys. {\bf 71}, S96 (1999).}

\bibitem{novaes} S. F. Novaes, ``Standard model: An introduction,'' 
\arXivOld{hep-ph/0001283}.

\bibitem{moreira} M. A. Moreira, ``O modelo padr\~ao da f\'isica de part\'iculas,'' 
\doi{10.1590/S1806-11172009000100006} {Revista Brasileira de Ensino de F\'isica {\bf 31},  1306  (2009).}

\bibitem{finite1} J. I. Kapusta and P. Landshoff, ``Finite-temperature field theory,'' 
\doi{} {J. Phys. G: Nucl. and Part. Phys. {\bf 15},  267 (1989).}

\bibitem{finite3} A. J. Niemi and G. W. Semenoff, ``Thermodynamic calculations in relativistic finite-temperature quantum field theories,'''
\doi{10.1016/0550-3213(84)90123-8} {Nucl. Phys. B {\bf 230}, 181 (1984).}

\bibitem{realandimaginary} N. P. Landsman and C. G. Van Weert, ``Real-and imaginary-time field theory at finite temperature and density,'' 
\doi{10.1016/0370-1573(87)90121-9} {Phys. Rep. {\bf 145}, 141 (1987).}

\bibitem{Matsubara} T. Matsubara, ``A New Approach to Quantum-Statistical Mechanics'',
\doi{10.1143/PTP.14.351} {Prog. Theor. Phys. {\bf 14}, 351 (1955).}

\bibitem{finite2} A. J. Niemi and G. W. Semenoff, ``Finite-temperature quantum field theory in minkowski space,''
\doi{10.1016/0003-4916(84)90082-4}  {Annals of Physics {\bf 152}, 105 (1984).}

\bibitem{Schwinger}J. Schwinger, ``Brownian Motion of a Quantum Oscillator,'' 
\doi{ 10.1063/1.1703727} {J. Math. Phys. {\bf 2}, 407 (1961).} 

\bibitem{tfd1} Y. Takahashi and H. Umezawa, ``Thermo field dynamics,'' 
Collective Phenomena {\bf 2}, 55 (1975); Reprinted in \doi{10.1142/S0217979296000817}{Int. J. Mod. Phys.B {\bf 10}, 1755 (1996)}.

\bibitem{khannatfd}  F.~C.~Khanna, A.~P.~C.~Malbouisson, J.~M.~C.~Malbouisson and A.~R.~Santana,
``Thermal quantum field theory - Algebraic aspects and applications,''
World Scientific Publishing Company (2009).

\bibitem{Umezawa:1982nv}
H.~Umezawa, H.~Matsumoto and M.~Tachiki,
``Thermofield Dynamics and Condensed States,''
North-Holland, Amsterdam (Netherlands) (1982).

\bibitem{Umezawa:1993yq}
H.~Umezawa,
``Advanced field theory: Micro, macro, and thermal physics,''
American Institute of Physics (1995).

\bibitem{lietfd} A. E. Santana and F. Khanna, ``Lie groups and thermal field theory,'' 
\doi{10.1016/0375-9601(95)00394-I} {Phys. Lett. A {\bf 203}, 68 (1995).}

\bibitem{experimental1}L. Myers, J. Annand, J. Brudvik, G. Feldman, K. Fissum, H. Grie{\ss}hammer, K. Hansen, S. Henshaw,
L. Isaksson, R. Jebali, et al., ``Measurement of compton scattering from the deuteron and an improved
extraction of the neutron electromagnetic polarizabilities,'' 
\doi{10.1103/PhysRevLett.113.262506} {Phys. Rev. Lett. {\bf 113},  262506 (2014).}

\bibitem{experimental2} C. Adloff, V. Andreev, B. Andrieu, T. Anthonis, V. Arkadov, A. Astvatsatourov, A. Babaev, J. B\"{a}hr,
P. Baranov, E. Barrelet, et al., ``Measurement of deeply virtual compton scattering at HERA,'' 
\doi{10.1016/S0370-2693(01)00939-X} {Phys. Lett. B {\bf 517}, 47 (2001).}

\bibitem{experimental3} D. Drechsel, B. Pasquini, and M. Vanderhaeghen, ``Dispersion relations in real and virtual Compton scattering,'' 
\doi{ 10.1016/S0370-1573(02)00636-1} {Phys. Rep. {\bf 378}, 99 (2003).}

\bibitem{experimental4} H. W. Grie{\ss}hammer, J. A. McGovern, D. R. Phillips, and G. Feldman, ``Using effective field theory to
analyse low-energy compton scattering data from protons and light nuclei,'' 
\doi{10.1016/j.ppnp.2012.04.003} {Prog. Part. Nucl. Phys. {\bf 67}, 841 (2012).}

\bibitem{comptonclosed-time} H.-H. Xu and C.-H. Xu, ``Compton scattering at finite temperature,''
\doi{10.1103/PhysRevD.52.6116} {Phys. Rev. D {\bf 52}, 6116 (1995).}

\bibitem{Das} A. Das, ``Finite temperature field theory,''
World Scientific (1997). 

\bibitem{Ahmed} M. A. A. Ahmed, H. Zainuddin and N. M. Shah, ``Real-time thermal self-energies: in the variational bases and spaces.''
\doi{10.1016/j.rinp.2022.105691} {Res. Phys. {\bf 39}, 105691 (2022).}

\bibitem{compton} D. Millar, ``A calculation of the differential cross section for Compton scattering in tree-level quantum electrodynamics,'' 
{Lecture notes (2014).}

\bibitem{ale1} A. F. Santos and F. C. Khanna, ``Quantized gravitoelectromagnetism theory at finite temperature,''
\doi{10.1142/S0217751X16501220} {Int. J. Mod. Phys. A {\bf 31},  1650122 (2016).}


\end{thebibliography}
\end{document}